\documentclass{ptapap}
\usepackage{amsmath}
\usepackage{amssymb}
\usepackage{color}

\usepackage{soul}

\author{Mohammad-Hassan Naddaf}[CFT, CAMK]
\author{Bo\.zena Czerny}[CFT]
\author{Ryszard Szczerba}[Torun]
\affil[CFT]{Center for Theoretical Physics,
  Lotnikow 32/46, 02--668 Warszawa, Poland}
\affil[CAMK]{Nicolaus Copernicus Astronomical Center,
  Bartycka 18, 00--716 Warszawa, Poland}
\affil[Torun]{Nicolaus Copernicus Astronomical Center,
  Rabianska 8, 87--100 Torun, Poland}
  
\title{R-L Relation in realistic FRADO Model}

\begin{document}

\maketitle

\begin{abstract}

In Failed Radiatively Accelerated Dusty Outflow (FRADO) model which provides the source of material above the accretion disk (AD) as an option to explain the formation mechanism of Broad Line Region (BLR) in AGNs, the BLR inner radius ($\rm{BLR}_{in}$ hereafter) is set by the condition that the dust evaporates immediately upon departure from the AD surface.
On the other hand, the location of BLR clouds obtained observationally via reverberation mapping shows some scaling with the source luminosity, so-called RL relation. We assume $\rm{BLR}_{in}$ to be the location of BLR clouds, then using a realistic expression for the radiation pressure of an AD, and having included the proper values of dust opacity, and shielding effect as well, we report our numerical results on calculation of $\rm{BLR}_{in}$ based on FRADO model. We investigate how it scales with monochromatic luminosity at 5100$\textup{\AA}$  for a grid of black hole masses and Eddington ratios to compare along with the FRADO analytically predicted RL directly to observational data.

\end{abstract}

\section{Introduction}

Active galaxies can be used to test cosmological models due to RL relation. BRL location can be determined observationally from the time delay of the broad line emission with respect to underlying continuum. The RL relation is usually calibrated for low redshift sources \citep{2013ApJ...767..149B}, assuming standard cosmology. However, FRADO model considering the fact that dust can exist in outer parts of AD \citep{2008MNRAS.383..581D}, allows to calculate RL relation directly from the theory \citep{2011A&A...525L...8C}. The 1-D analytical form of the model has been developed \citep{2017ApJ...846..154C}. In the present numerical work, we aim at a realistic 3-D picture of the model to compare to 1-D analytical model and observational data.

\section{$\rm{BLR}_{in}$ in FRADO Model}

The analytical model yields the $\rm{BLR}_{in}$ simply by setting the effective temperature in Shakura-Sunyaev (S-S) AD equal to the dust sublimation temperature $T_{\mathrm{sub}}$
\begin{equation}
\mathrm{BLR}_{\mathrm{in}} =
\left( \frac{3 G M_{\mathrm{BH}}\ \dot{m}\ \dot{M}_{\mathrm{Edd}}}{
8 \pi \sigma_B\ T^4_{\mathrm{sub}}} \right)^{1/3},
\end{equation}
where $M_{\mathrm{BH}}$, $\dot{m}$, and $\dot{M}_{\mathrm{Edd}}$ are black hole mass, dimensionless accretion rate, and Eddington accretion rate, respectively. The monochromatic luminosity of AD at 5100$\textup{\AA}$ for low mass sources \citep{2011A&A...525L...8C}, is well approximated as
\begin{equation}
\log \left(\frac{L_{5100}}{10^{44}\ \mathrm{erg/s}}\right) =
\frac{2}{3} \log (M_{\mathrm{BH}}\ \dot{m}\ \dot{M}_{\mathrm{Edd}})
+ \log \cos i  - 39.882,
\end{equation}
where $i$ is the viewing angle. Combining this equation with Eq. 1, one can find an interesting analytical RL relation which is independent of $M_{\mathrm{BH}}$, $\dot{m}$, and $\dot{M}_{\mathrm{Edd}}$
\begin{equation}
\log \left(\frac{\mathrm{BLR}_{\mathrm{in}}}{\mathrm{lt.day}}\right) =
\frac{1}{2}\ \log \left(\frac{L_{5100}}{10^{44}\ \mathrm{erg/s}}\right)
- \frac{4}{3} \log T_{\mathrm{sub}}
- \frac{1}{2}\ \log \cos i
+ 5.2436.
\end{equation}

In 3-D picture, unlike the 1-D model we consider that the radiation acting on dust is coming from all AD radii. However, it can be argued that the radiation from AD is partly shielded due to either inner failed Compton-driven wind or cold material forming at the edge between the inner hot ADAF and a cold disk; otherwise too intense radiation from central region of AD causes too early dust evaporation within a large range of AD radii. So, we incorporate the effect of shielding into the computations via our Patch-Model in which we assume that the radiation only comes from a small patch of the disk whose radius $s$ is linearly proportional to the actual height $H$ of the clump from the disk surface, i.e. $s=\alpha H$. The patch is always centered at a radius where the clump is flying above, and increases in size as the cloud goes up. The energy absorbed by dust in our model is given by
\begin{equation}
Q_{\mathrm{abs}} =
\int_{\lambda_i}^{\lambda_f}
\int_{\mathrm{patch}} f\ (\textbf{r}, M_{\mathrm{BH}}, \dot{m},
K_{\mathrm{abs}}(\lambda), T_{\mathrm{eff}}(R), R, \varphi, \lambda, C)\
d\textbf{a}\ d\lambda,
\end{equation}
where $\textbf{r}$ is the position vector of the clump in cylindrical coordinates in which the black hole is at the center, and $z=0$ corresponds to equatorial plane of AD, $K_{\mathrm{abs}}$ is dust absorption opacity as a function of wavelength,  $T_{\mathrm{eff}}(R)$ is the effective temperature of S-S AD as a function of radius, where the AD is assumed to radiate locally at a given radius as a black body, $R$ and $\varphi$ indicate the location of infinitesimal surface areas of AD in polar coordinates, and $C$ denotes physical constants. We assume that dust instantly emits the absorbed irradiated energy in the form of black body radiation. Once $Q_{\mathrm{abs}}$ exceeds $Q_{\mathrm{emit}}(T_{\mathrm{sub}})$, the dust gets immediately sublimated.

\section{Numerical Setup \& Results}

We consider the dust content of BLR to follow the MRN dust model proposed by \citet*{1977ApJ...217..425M}. It consists of silicate and graphite grains, for which the mean absorption cross-section per dust mass (equivalent to dust opacity) is computed using MCDRT \citep{1997A&A...317..859S} and KOSMA--$\tau$ PDR \citep{2013A&A...549A..85R} codes. Dust sublimation temperature is set to be 1500 K. The proportionality constant $\alpha$ in Patch-Model is taken to be $1.5$. A range of $M_{\mathrm{BH}} = 10^7, 10^8, 10^9 M_{\odot}$, and $\dot{m} = 0.01, 0.1, 1$ (for accretion efficiency of $0.1$) is chosen as the model grid.

For each pair of $M_{\mathrm{BH}}$, $\dot{m}$ in our grid we computed the sublimation geometrical location above which dust cannot survive irradiation. Having the disk thickness computed using a numerical code for vertical structure of radiation dominated AD \citep{1999MNRAS.305..481R,2016ApJ...832...15C}, which includes all opacities important at large disk radii, we find the crossing radius of the sublimation surface with the disk surface to be $\mathrm{BLR}_{\mathrm{in}}$.

\begin{figure}
\includegraphics[width=\textwidth]{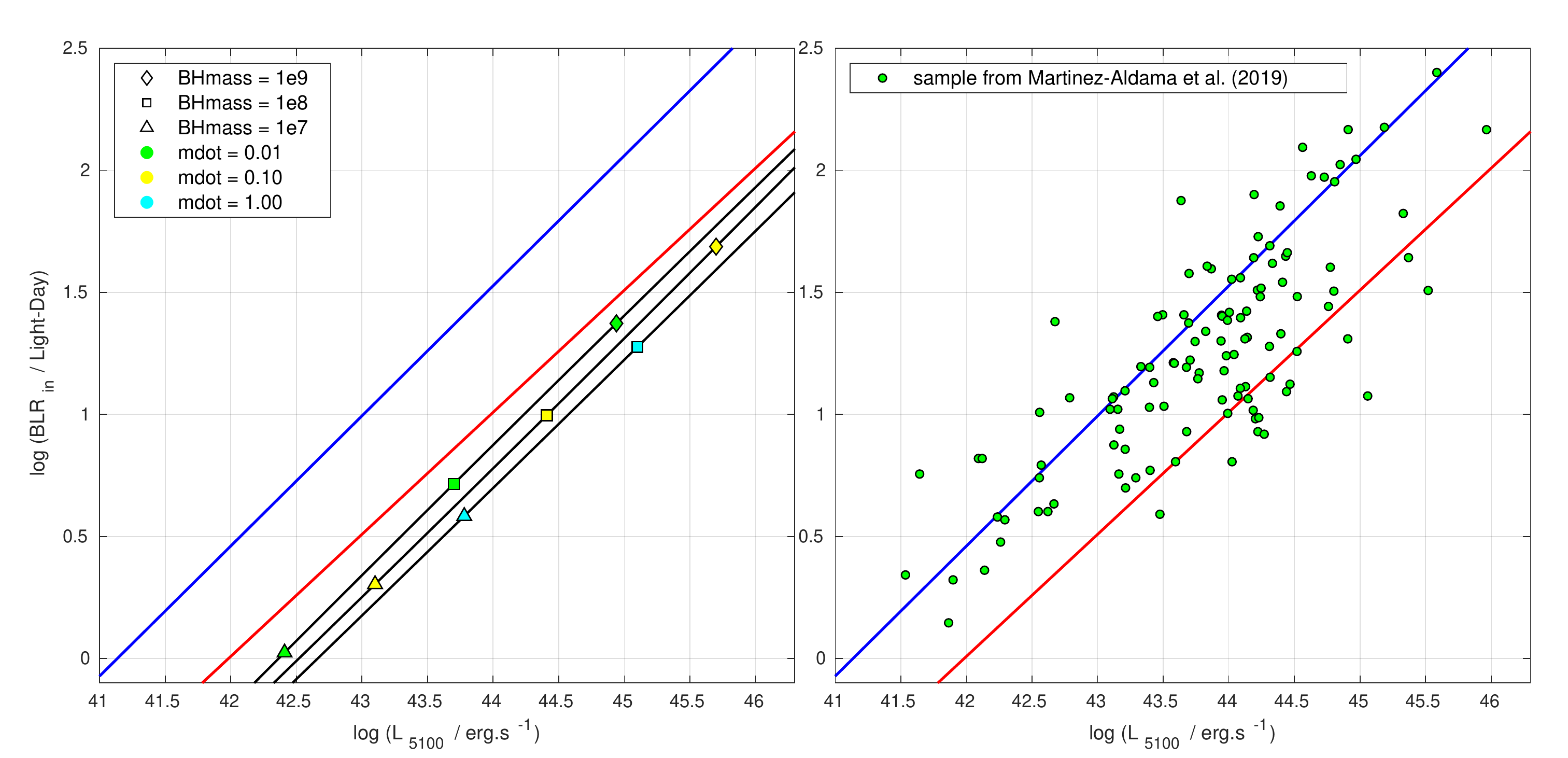}
\caption{RL relations per Eddington ratios from realistic FRADO model shown by black solid lines (left) vs. observational sample (right). The blue and red solid lines represent the \citet{2013ApJ...767..149B} RL and analytical FRADO RL, respectively.}
\label{fig:motion}
\end{figure}

Finally, setting the viewing angle to be zero (faced-on disk) for simplicity, the results are plotted against the numerically computed values of $L_{5100}$ \citep{2003A&A...412..317C} in the left panel of Fig. \ref{fig:motion}. A sample of observational data \citep{2019ApJ...883..170M} is provided in the right panel of Fig. \ref{fig:motion} for comparison.

\section{Conclusion}

RL relations based on $\mathrm{BLR}_{\mathrm{in}}$ predicted by realistic FRADO model are located below the \citet{2013ApJ...767..149B} RL relation which is for low redshift sources. However, they interestingly follow the same slope as \citet{2013ApJ...767..149B}, they look to be more consistent with the recent BLR size measurements for high Eddington ratio sources. Moreover, the model predicts that the real BLR is extended, so the realistic time delays from the entire BLR will be longer. In the future we will focus on the dynamics of clouds in BLR more in detail to find the BLR geometry and range of possible time delays.

\acknowledgements{The project was partially supported by NCN grant No. 2017/26/A/ST9/ 00756 (Maestro 9), and by MNiSW grant No. DIR/WK/2018/12.}

\bibliographystyle{ptapap}
\bibliography{naddaf}

\end{document}